# Stationary Multi-source AI-powered Real-time Tomography (SMART)

Weiwen Wu[1,2 #], Yaohui Tang[3#], Tianling Lv[4], Wenxiang Cong[1], Chuang Niu[1], Cheng Wang[3], Yiyan Guo[3], Peiqian Chen[5], Yunheng Chang[4], Ge Wang[1*], Yan Xi[4*]

[1]Biomedical Imaging Center, Center for Biotechnology and Interdisciplinary Studies, Department of Biomedical Engineering, Rensselaer Polytechnic Institute, Troy, NY, USA

[2]School of Biomedical Engineering, Sun Yat-Sen University (SYSU), Gongchang Road, Guangming District, Shenzhen, Guangdong, 518100, China

[3]Med-X Research Institute, School of Biomedical Engineering, Shanghai Jiao Tong University, 1954 Hua Shan Road, Shanghai, 200030, China.

[4]Jiangsu First-Imaging Medical Equipment Co., Ltd., Jiangsu, 226100, China.

[5]Tongren Hospital, Shanghai Jiao Tong University School of Medicine, 1111 Xianxia Road, Shanghai, 200336, China

# Co-first authors, and * Co-corresponding authors

**Abstract:** Over the past decades, the development of CT technologies has been largely driven by the needs for cardiac imaging but the temporal resolution remains insufficient for clinical CT in difficult cases and rather challenging for preclinical micro-CT since small animals, as human cardiac disease models, have much higher heart rates than human. To address this challenge, here we report a Stationary Multi-source AI-based Real-time Tomography (SMART) micro-CT system. This unique scanner is featured by 29 source-detector pairs fixed on a circular track to collect x-ray signals in parallel, enabling instantaneous tomography in principle. Given the multi-source architecture, the field-of-view only covers a cardiac region. To solve this "interior problem", an AI-empowered interior tomography approach is developed to synergize sparsity-based regularization and learning-based reconstruction. To demonstrate the performance and utilities of the SMART system, extensive results are obtained in physical phantom experiments and animal studies, including dead and live rats as well as live rabbits. The reconstructed volumetric images convincingly demonstrate the merits of the SMART system using the AI-empowered interior tomography approach, enabling cardiac micro-CT with the unprecedented temporal resolution of 30ms, which is an order of magnitude higher than the state of the art.

I. Introduction

Since the first commercial computed tomography (CT) technology was introduced that allowed electrocardiogram (ECG)-gated CT imaging (Electron Beam CT, EBCT) in the early 1990s [1], cardiac CT has become an indispensable tool for diagnosis and treatment of cardiovascular diseases. The current multi-row detector CT (including dual source CT, DSCT) helps improve image quality and diagnostic performance for many patients. However, patients often have special conditions such as severe arrhythmia, atrioventricular block, etc., are still contraindications to cardiac CT. Improving temporal resolution further is highly desirable to reduce cardiac motion artifacts. In addition to anatomical cardiac CT, CT perfusion (CTP) is a functional imaging technique that relies on high temporal resolution. CTP is mainly used to diagnose cerebral ischemia and tumor heterogeneity [2]. After rapid intravenous injection of contrast agent, continuous CT scanning is performed on a selected slice of interest to monitor local blood perfusion and computes physiological parameters, such as mean transit time (MTT), permeability–surface area product (PS), blood flow, and blood volume. The quality and accuracy of MTT and PS depend on the imaging speed at which an adequate projection dataset is acquired.

CT equipment with extremely high temporal resolution is not only valuable for diagnosis of the human cardiac diseases but also in preclinical research. Mice, rabbits, dogs, etc. are commonly used in preclinical research. Their heart rates (say, 600 beats per minute for a rat) are much higher than that of the human, and make imaging of the heart and lungs very challenging. Here we report our recently prototyped first-of-its-kind micro-CT scanner to image heart and lungs of rats and rabbits at the highest imaging speed so far.

As a common non-invasive imaging tool, a CT scanner usually contains only one or two source–detector assemblies with a sub-optimal temporal resolution [4]. Over the past decades, major temporal resolution improvements are being made with an increasingly faster rotation speed [5], two tube-detector pairs [6], and advanced reconstruction techniques [7]. Commonly, a CT scanner with a single x-ray source scans at a speed as fast as up to 3 Hz. Ultimately, the centrifugal force limits the scanning speed. Although the rotating CT gantry dominates in hospitals and clinics, it fails to provide an ideal imaging performance in difficult cases [8].

Worldwide, cardiovascular diseases (CVDs) are the leading cause of death, taking almost 17.9 million lives annually [9]. CVDs includes a group of disorders of the heart and associated vasculature, such as coronary heart disease, cerebrovascular disease, rheumatic heart disease, and other disorders. Four out of five CVD deaths are due to heart attacks and strokes. Dynamic cardiac CT studies demand better technologies and has been a primary driving force for development of the CT field. Since CT temporal resolution is not sufficiently high, ECG-gating is widely employed to account for a quasi cyclical cardiac motion, improving temporal resolution and minimizing image artifacts. Unfortunately, this approach has major limitations that become most evident in patients with irregular and/or fast heart rates. Furthermore, radiation exposure is high with ECG-gated cardiac CT, given the requirement for continuous overlapped scanning and retrospective data grouping.

Extensive efforts have been made to address these challenges. The system with multiple source-detector chains is a feasible solution. To reach this goal, various system designs were proposed. The rationale is that increasing the number of source-detector pairs in a gantry will reduce the data acquisition time and improve the temporal resolution [10]. The first example is the multisource CT prototype known as the dynamic spatial reconstructor (DSR) [11], which still demands a mechanical scan. Subsequently, several multi-source CT schemes were designed. Liu et al. demonstrated the improved image quality in a simulated five-source cone-beam micro-CT using a Feldkamp-type reconstruction algorithm [12]. Zhao et al. conceptualized a triple-source helical/saddle cone-beam CT system and developed an exact volumetric reconstruction algorithm. Cao et al. proposed an 80 multi-source interior CT architecture that employs three stationary x-ray source arrays and three detector operated in the interior tomography mode [13]. For these multiple source-based x-ray imaging system designs, a general challenge is how to collect high-quality data and perform interior CT reconstruction [14]. For interior tomography, x-ray beams are restricted to travel through a local region of interest (ROI), with the measurement on the ROI compromised by both surrounding tissues, Poisson noise and Compton scattering [15,16]. Nevertheless, interior tomography enables utilization of smaller x-ray detectors, takes more projections at the same time, and provides ultrahigh temporal resolution. Clearly, the multi-source interior CT architecture has the potential to achieve all-phase cardiac CT imaging. However, these multiple-source CT cardiac imaging systems have not been prototyped so far because of the complexity of the system engineering, data processing and image reconstruction, as well as cost incurred in such a highly non-trivial undertaking.

Over recent years, the use of artificial intelligence (AI) [17], specifically deep learning, has become instrumental in the medical imaging field [18]. For dynamic cardiac imaging, the adaption of deep learning techniques is now the mainstream to remove image artifacts in the cases of limited and compromised measurements [19]. Bello et al. took image sequences of the heart acquired using cardiac MRI to create a time-resolved segmented dynamic volume using a fully convolutional network aided by anatomical shape prior [20]. To provide a high-quality image in phase-contrast magnetic resonance imaging, Vishnevskiy et al. proposed an efficient model-based deep neural reconstruction network to avoid hyperparametric turning and expensive computational overhead of compressed sensing reconstruction methods for clinical aortic flow analysis [21]. To improve the overall quality of 4D CBCT images, two CNN models, named N-Net and CycN-Net, were proposed in [22].

The state-of-the-art cardiac CT scanner with a wide-area detector covers an entire heart within a single cardiac cycle. For example, the Revolution$^{TM}$ CT scanner (GE HealthCare) achieves a temporal resolution 140ms [23]. As the posterior left ventricular wall moves at a maximum velocity of 52.5 mm/s, a scan time of 19.1ms or less is ideal to avoid motion artifacts. If the average mean velocity is considered, the scan time should still be 41.8ms. Currently, even if deep learning is used with the current CT systems, the temporal resolution remains sub-optimal. It is feasibility to incorporating the deep learning into the multi-source imaging system. Due to the restricted field-of-view (FOV) of the multiple source-detector CT system, one typically encounters the interior CT reconstruction problem.

In this study, we first prototype a Stationary Multi-source AI-based Real-time Tomography (SMART) system, where 29 source-detector pairs are fixed on a circular track to collect x-ray signals in parallel. The new system architecture leads to a major improvement in temporal resolution. To achieve the high-quality dynamic images from truncated and sparse measurements, we developed the AI-empowered interior tomography approach  network, where we utilize the relationship among different timeframes and reconstruct raw time-group average images. Then, the interior CT reconstruction network is trained to recover high-quality deep group-average images from raw time-group average images. Also, the sparsity dictionary was trained from prior deep group-average images and further reconstruct high-quality group-average image using a compressed sensing model. Finally, we obtain the final reconstruction depending on the current measurement and the high-quality group-average prior image. Our SMART imaging system has the highest high temporal resolution so that we can obtain high-quality interior CT images of the heart and lungs in small animal models.

The rest of the paper is organized as follows. In the next section, we introduce our multi-source CT system prototype, the first-of-its-kind Stationary Multi-source AI-based Real-time Tomography (SMART) system for dynamic cardiac imaging. Then, we describe our deep reconstruction method that integrates temporal prior, sparsified prior and deep network prior into our AI-empowered interior tomography approach. In the third section, we report our experimental results, showing the feasibility and merits of our SMART system and AI-empowered interior tomography method. In the last section, we discuss related issues and conclude the paper.

## II. Results

### A. SMART System Prototype

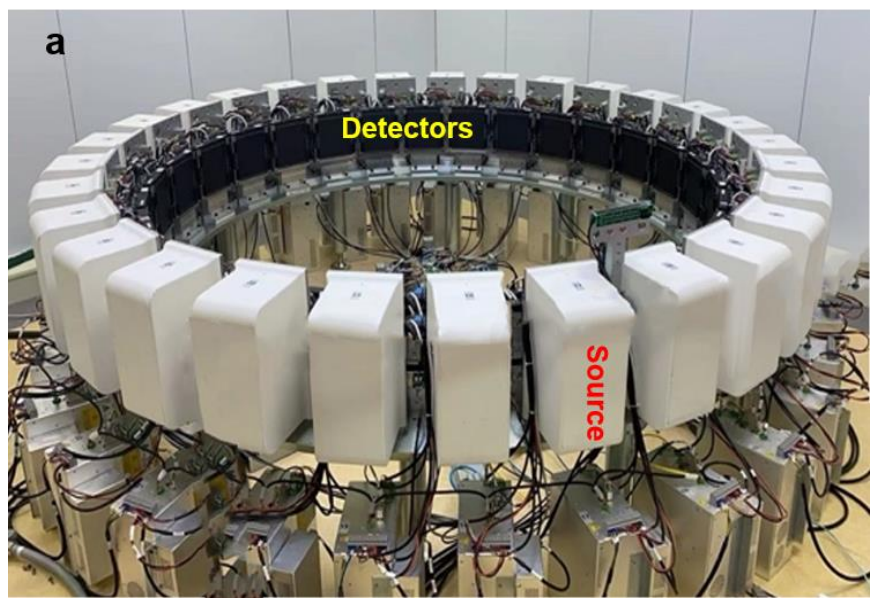

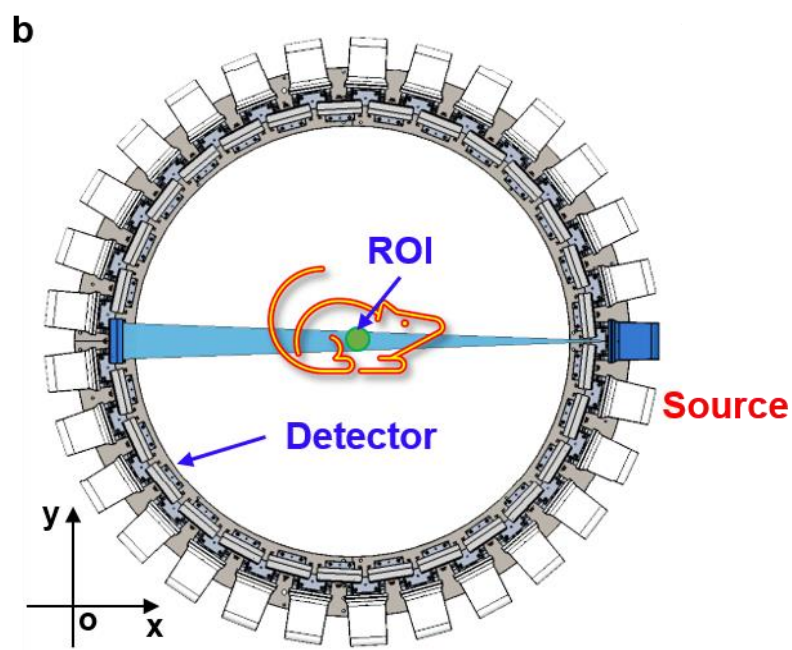

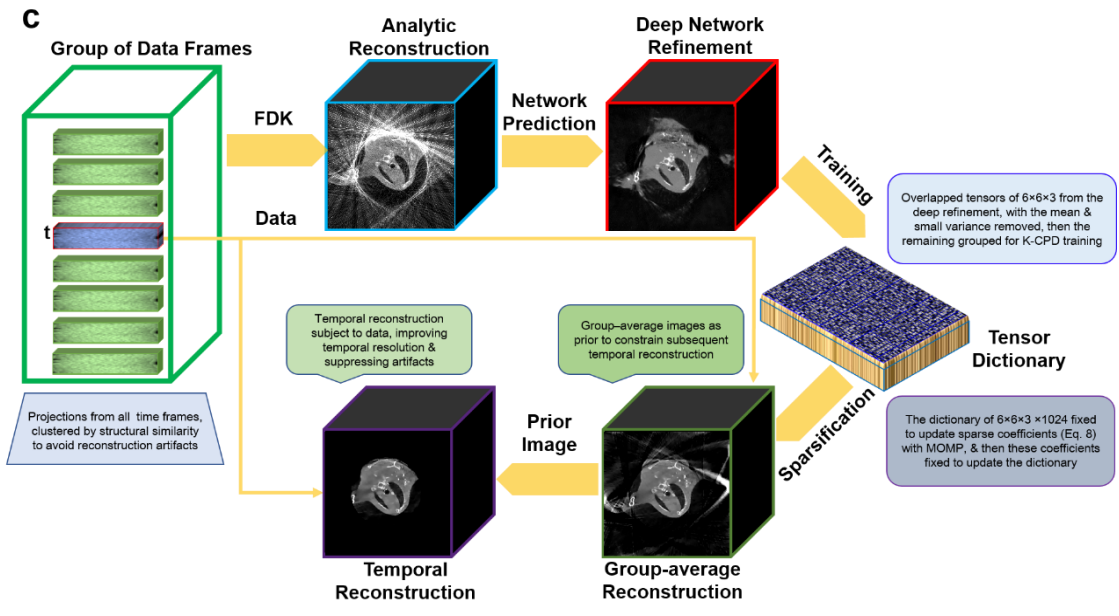

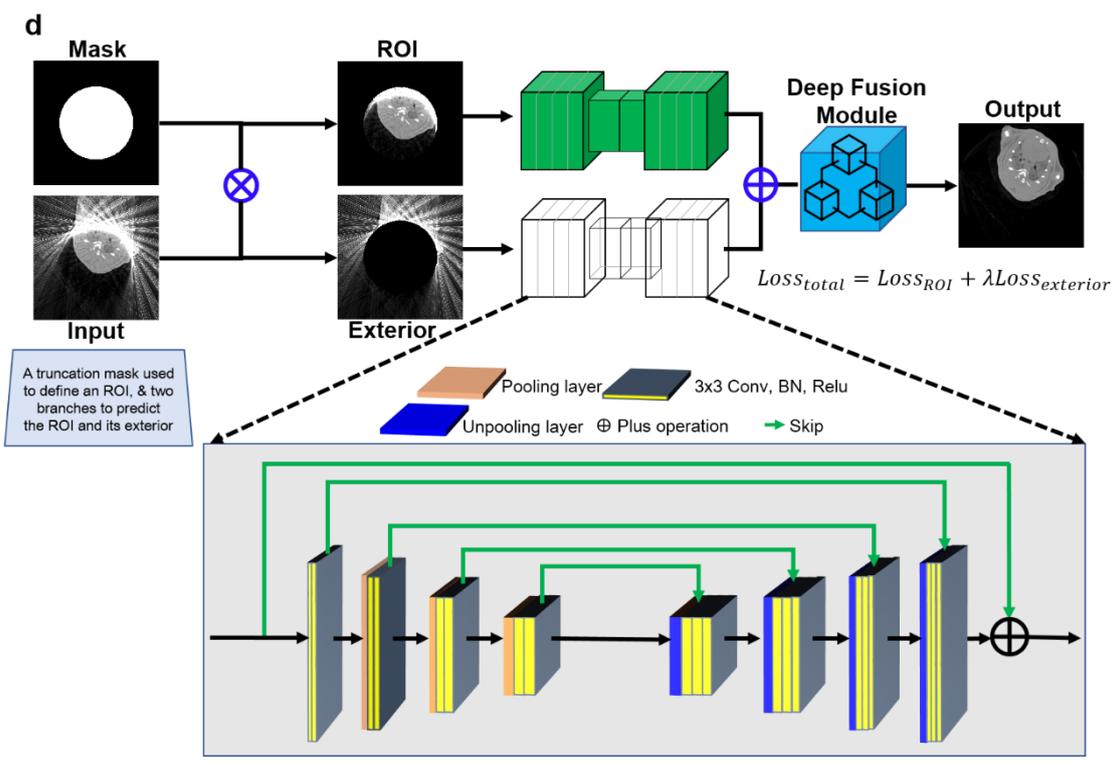

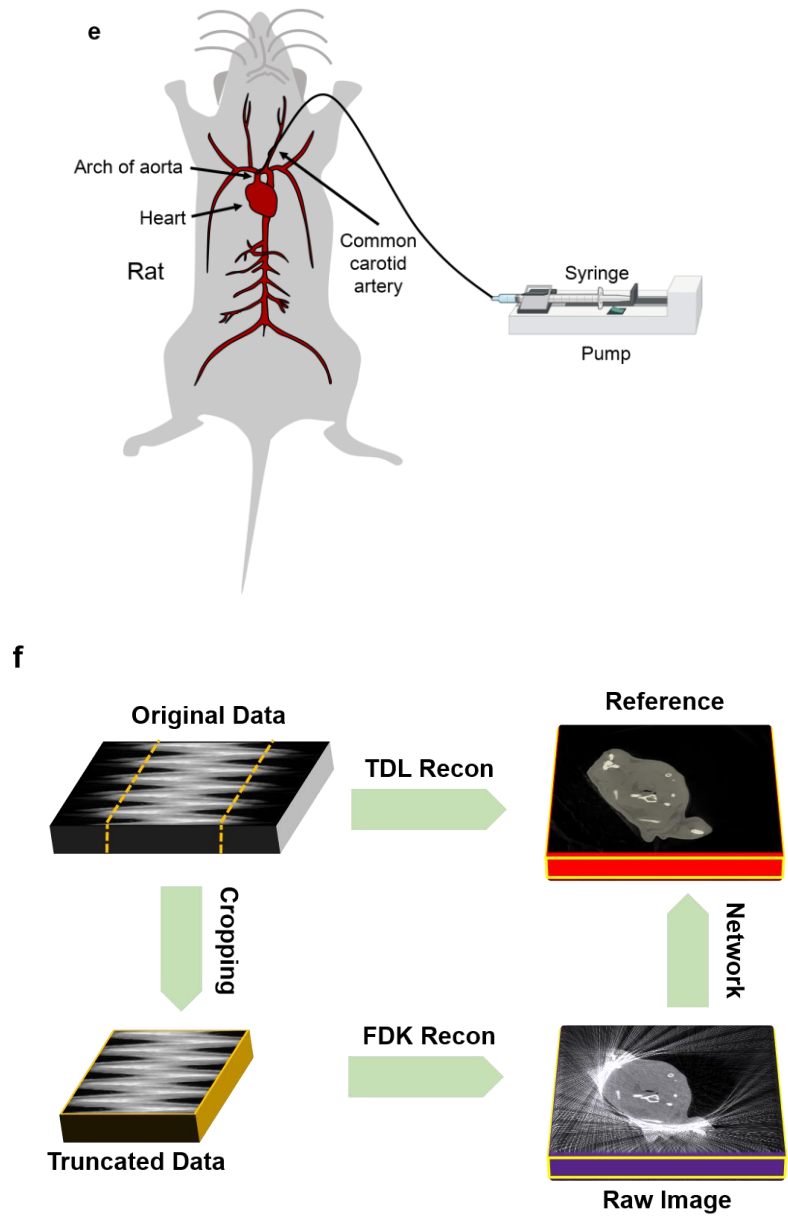

Figure 1. Multi-source "SMART" preclinical CT prototype and reconstruction frameowrk for dyamic image reconstruction. (a) A photograph of the real system; (b) the imaging geometry; (c) the flowchart of the SMART reconstruction framework; (d) the interior CT network in the first SMART reconstruction stage; (e) animal model for contrast injection with dynamic imaging; (f) training, validation and testing for deep neural network-based interior reconstruction.

The SMART system consists of 29 x-ray source and detector pairs, all of which are fixed on a circular track. There are a 5 kW monoblock x-ray source and an IGZO flat panel detecor of $153.6 \times 153.6$ mm$^2$ imaging area in each pair. The source-isocenter-distance (SID) and detector-isocenter-distance (DID) are set to 2,000 mm and 1,000 mm, respectively. Each detector cell covers an area of $0.2 \times 0.2$ mm$^2$. The x-ray beam generated by the x-ray source is collimated through the gap between neighboring detectors. An animal to be imaged is placed inside the imaging ring with a zooming factor of 1.87 [24].

During the data collection, these imaging pairs are simulaneously turned on to capture cone-beam projections. A sequence of x-ray pulses is fired at 10 frames per second (fps). Since the x-ray sources are symmetrically distributed, a rotation range of 12.4 degrees is sufficient for high density sampling, which can be used for evaluation of the imaging fidelity. In rat experiments, the x-ray energy is set at 70 kV, the current to 30 mA, and the pulse width of 20 ms. Since there is no grid mounted on the detector surface, projection calibration and scattering correction are done using our imaging software.

To obtain a high-quality prior image, we can collect sufficient data from many different timeframes to reconstruct a prior image. To the first order approximation, the precision rotation table is rotated to acquire data for different cardiac phases. These resultant projections from different timeframes can be considered as complete (sufficiently many viewing angles) but inconsistent due to cardiac motion. Then, we pre-process the projections and perform deep learning-based reconstruction. That is, we rearrange the projections in a chronological order for spatiotemporal sparsity-promoting image reconstruction in AI-empowered interior tomography approach. The overall workflow of our reconstruction approach is illustrated in Figure 1c, which can be divided into the three stages: deep group-prior prediction (shown in Figure 1d), sparsity group-prior reconstruction, and temporal image refinement. Figure 1e demonstrates the animal model with contrast injection to assess the performance of SMART in dynamic imaging. To obtain training datasets for deep reconstruction, here we design a data-generation process as shown in Figure 1f. More details about AI-empowered interior tomography reconstruction algorithm can be found in the **Method** section. In the following sections, we will assess the performance of AI-empowered interior tomography reconstruction network techniques with dead and alive rats as well as alive rabbits.

**B. Performance Evaluation on Dead Rats**

To demonstrate the feasibility of our proposed SMART system in cardiac imaging as well as the advantages of the reconstruction network, we first assessed the performance on dead rats. Figure 2 demonstrates volumetric reconstruction and rendering from 29 projections using different reconstruction algorithms. It is observed based on these results that our proposed approach can produce the best image quality with the least feature compromisation. Especially, our AI-empowered interior tomography approach results clearly demonstrate small blood vessels in the lungs. To further reveal the advantages of our reconstruction method, Figure 3 shows representative transverse slides reconstructed from 29 projections. It can be seen in Figure 4 that our approach produced the best reconstructed image. Specifically, the reconstructed results using the FDK algorithm contains the data trucation and sparse-view artifacts, blurring image details and edges. Compared with the FDK reconstruction, the SOTA compressed sensing approach offered a better reconstruction performance by removing the sparse-view and data-transverse artifacts but still failing to obtain clear edges and fine structures. On the other hand, our proposed network not only removed truncation and sparse-view artifacts but also recovered fine structures and edges faithfully. To quantitatively compare the results obtained using different reconstruction methods, the peak signal-to-noise ratio (PSNR) and structural similarity (SSIM) were employed, as also listed in Figure 3. Furthermore, the coronal and siggital reconstructed slices are also given in Figure 3, along with the quantitative results. From these results, it is clear that our approach reconstructed the images of high quality.

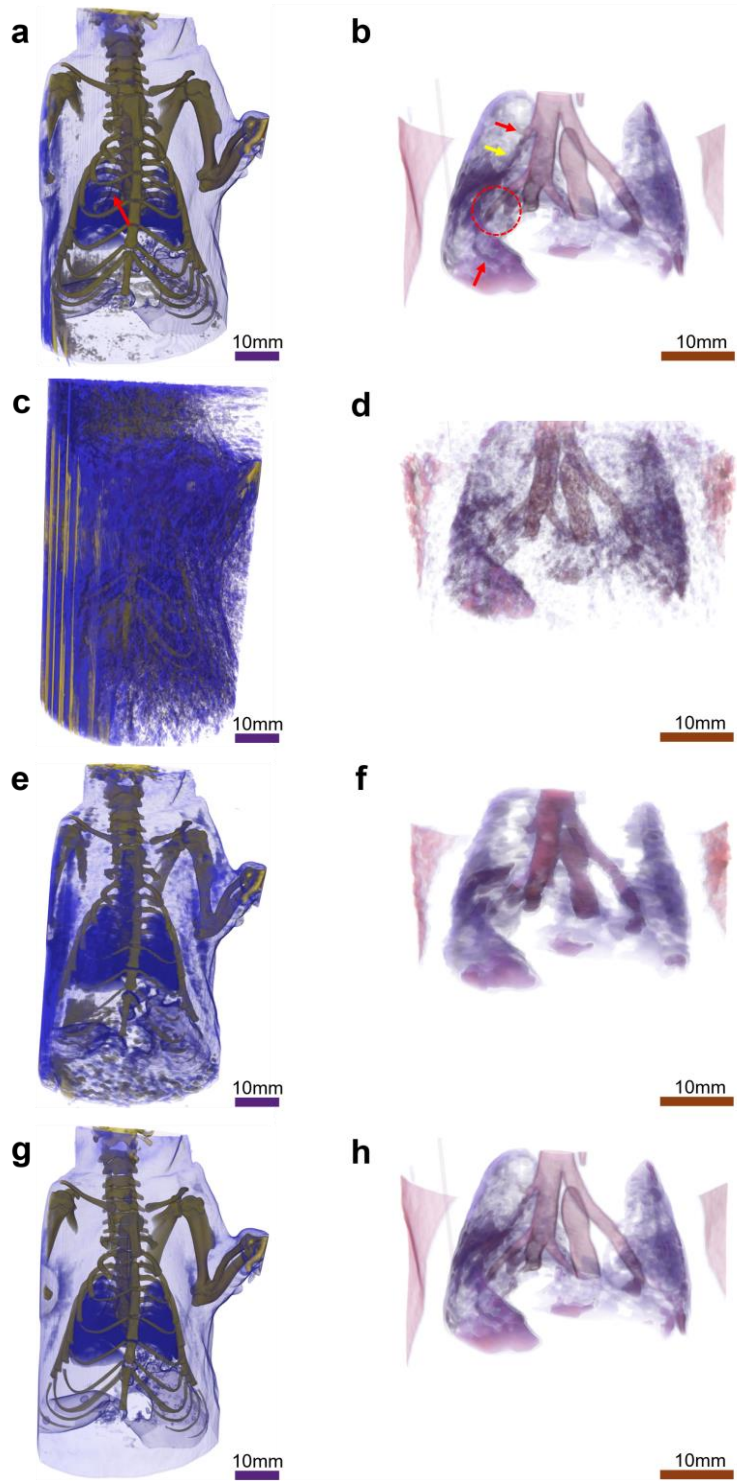

Figure 2. Volumetric rendering of an extracted lung region-of-interest using different reconstruction algorithms. (a) and (b) The reference from a full projection dataset without truncation using the TDL algorithm, (c) and (d), (e) and (f), (g) and (h) are the images reconstructed from only 29 views using FDK, SOTA and our AI-empowered interior tomography methods respectively. (b), (d), (f) and (h) represent the extracted lung region to further demonstrate AI-empowered interior tomography approach advantages.

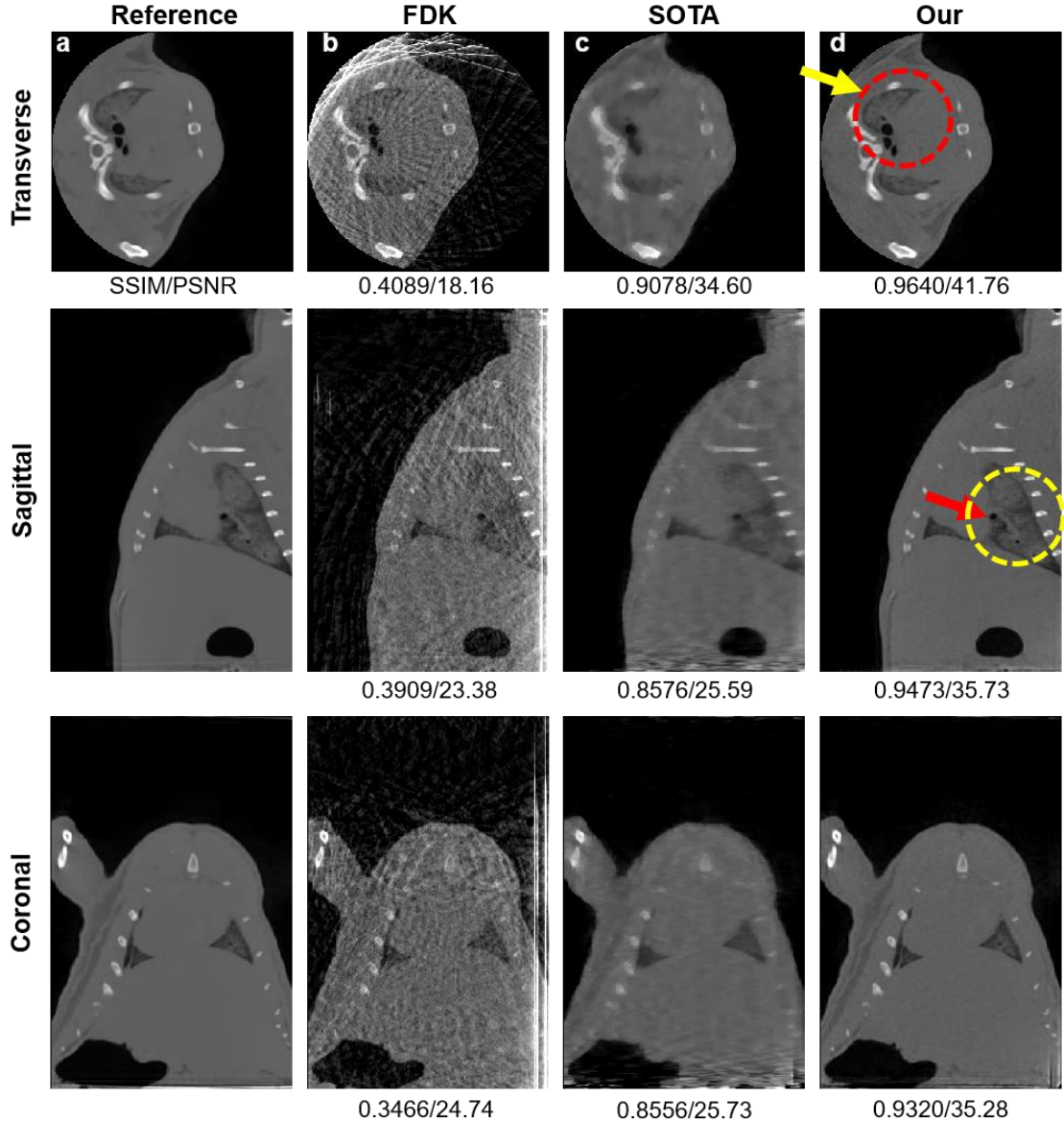

Figure 3. Representative transverse slices through a dead rat to validate our AI-empowered interior tomography reconstruction network. (a) The reference from a full projection dataset without truncation using the TDL algorithm. (b)-(d) The images reconstructed from only 29 views using FDK, SOTA and our methods respectively. The images (b), (c) and (d) from sparse data using different methods demonstrate remarkable image quality variations. The images (b) column is unacceptable due to strong artifacts, poor texture and inability to assess some anatomical structures. The images (c) column have much-reduced artifacts but look overly smooth, compromising texture for clinical usability. The images of (d) column using our approach has optimal image quality in terms of texture conservation (yellow circle), artifacts suppression, and clear visualization of small structures (red circle). The display window is [0 0.065] in terms of the linear attenuation coefficient.

## C. Dynamic Cardiac Imaging on Alive Rats

To show a dynamic cardiac imaging capability of the SMART system, typical reconstruction results from an alive rat are presented in Figure 4. It can be seen in Figure 4 that our AI-empowered interior tomography network provided dynamic cardiac features with well-defined edges. Since all source-detector pairs collected data simultaneously, the temporal resolution is 30ms. Specifically, cardiac features within 3D volume rendering clearly indicated by the small blood vessels are easily visualized in our method reconstructed images, while they are blurry in the SOTA reconstructions and even completely disappeared in the FDK results. More importantly, the bone structures and details are missing in both SOTA and FDK results, these features are found in our AI-empowered interior tomography approach results with 3D volume rendering images, Furthermore, the image structures within the ROI (indicated by the circle and arrows) in Figure 4 show that our proposed method satisfies the requirement of dynamic cardiac imaging, while the results obtained using the compressed sensing-based reconstruction method failed to do so. Exemplary coronal and sagittal slices from the alive rat are given in Figure 4 to further reveal the advantages of our proposed method.

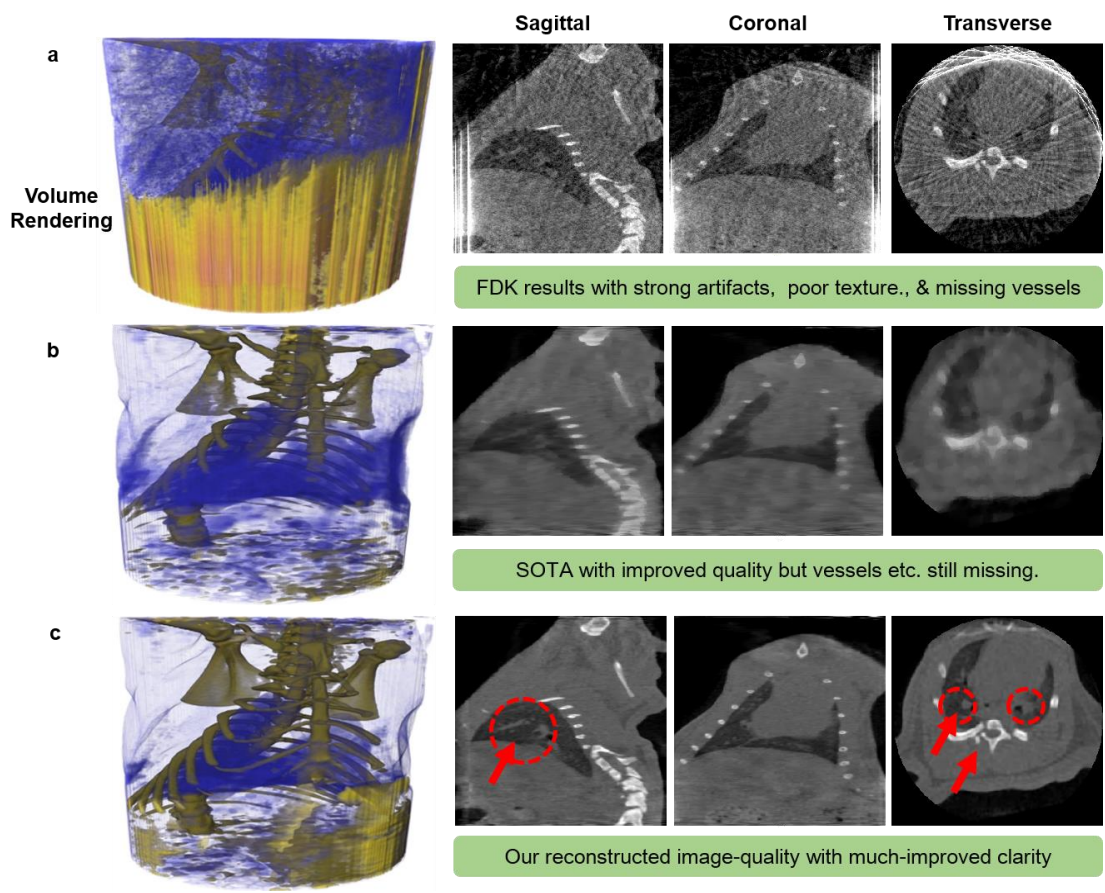

Figure 4. Reconstruction results of the alive rat to evaluate our reconstruction technique. (a)-(c) contain three images from transverse, coronal and sagittal reconstructed from only 29 views using FDK, SOTA, and our methods, respectively. The display window is [0 0.065] in terms of the linear attenuation coefficient. The images a with FDK reconstruction are unacceptable for strong artifacts, poor texture and undermined structures. The image from (b) are significantly better but they are overly smooth and quite blocky. The image of (c) reconstructed with our method have optimal image quality. The display window is [0 0.065] in terms of the linear attenuation coefficient.

To further demonstrate the advantages of the AI-empowered interior tomography method in the alive rat imaging study, the reconstruction results from three different timeframes are given in Figure 5. Our SMART results provide dynamic features with more faithful edges than the competing results. Additionally, our reconstruction algorithm provided finer blood vessels and it also can reconstruct the vessels changes with time. These cases show consistently that our AI-empowered interior tomography approach has a superior performance over competitors.

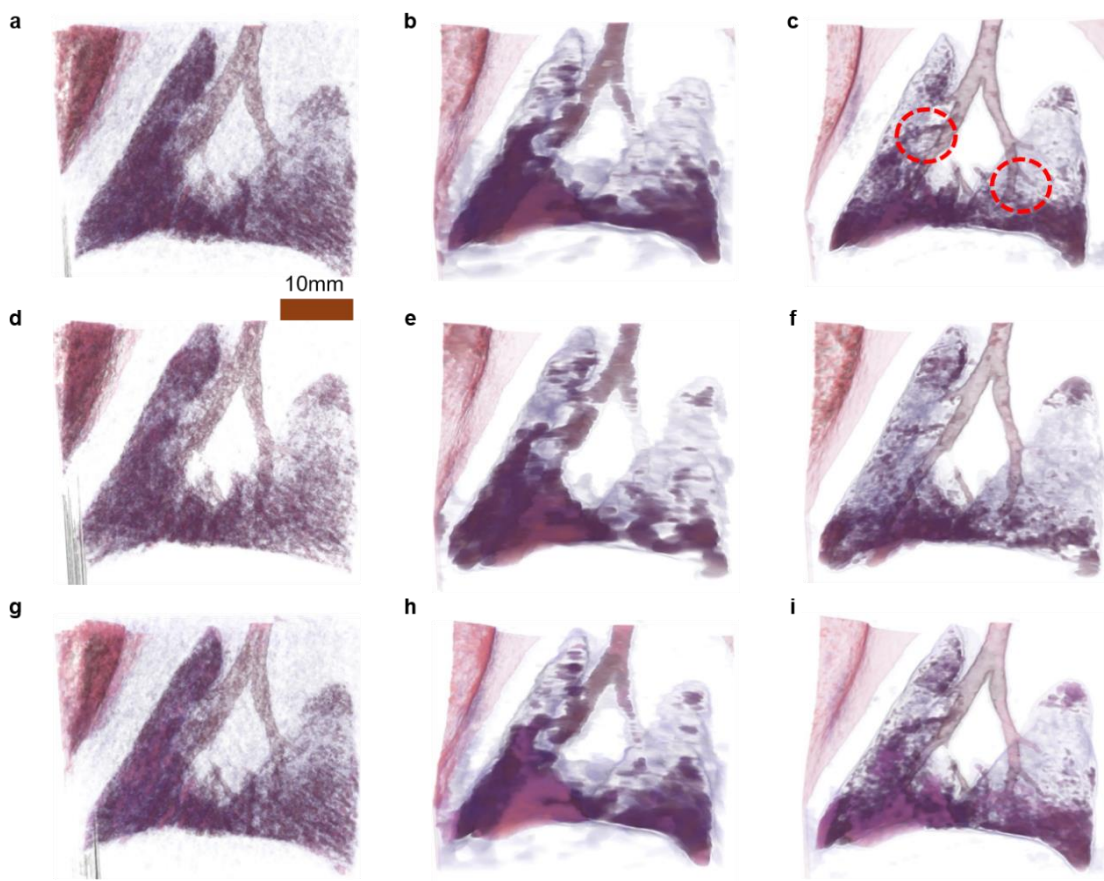

Figure 5. Sequential CT images of the alive rat in three timeframes to visualize dynamic changes with extracted tissue. The images in (a-c), (d-f) and (g-i) were reconstructed using the FDK, SOTA, AI-empowered interior tomography methods, respectively. The display window is [0 0.065] in terms of the linear attenuation coefficient.

### D. Dynamic Cardiac Imaging on Alive Rabbits

To fully test our SMART system on its interior tomographic imaging capability, we collected truncated projection datasets from an alive rabbit. The diameter of the rabbit is 140mm, well beyond the FOV size. In this study, the trained interior CT network using dead rats was employed. Figure 6 shows the reconstructed images from the alive rabbit with temporal resolution 30ms to evaluate image quality in terms of anatomical features. Compared with the FDK results, SOTA improved image quality regularized by the sparsity prior. However, SOTA oversmoothed details and edges, with some features lost and severe blocky artifacts introduced. In contrast, our method improved image quality by incorporating both deep prior learning and sparsity regularization. Specifically, the image feature indicated by the yellow circle is well preserved in our results, while it is difficult to observe in the SOTA reconstruction. As another example, the image structures indicated by the red circle in Figure 6 were corrupted to different degrees by competing methods, due to limited-angle, data truncation and sparse-view artifacts. In these cases, the imaging performance of our proposed approach is consistently better than the competitors.

To further demonstrate the advantages of our approach for dynamic imaging, the reconstructed results from different timeframes are given in Figure 7. Our approach provided more dynamic information with clearer image features than the competing methods. Again, these cases show that the imaging performance of our approach is superior to the competitors, and has a great potential for dynamic cardiac CT. It can be observed that our proposed method clearly observed the small vessels changes over other competitors, this point can be found within ROI indicated with red circle. To further demonstrate the advantages of our approach for dynamic imaging, the profiles along the blue position in Figure 7c were highlighted in Figure 7d, which clearly demonstrates the dynamic changes of the blood vessels and tissues. Furthermore, the reconstructed results from two different timeframes are given and magnified in Figure 8, where we compare the 2D projection and 3D reconstruction results.

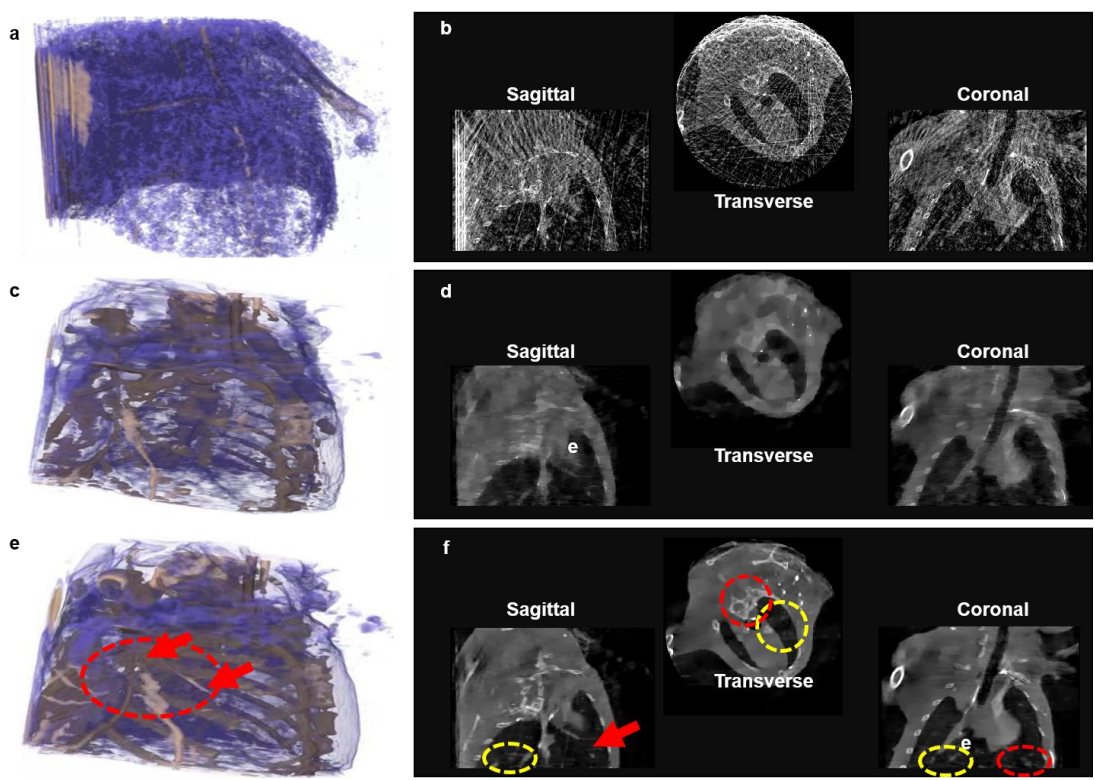

Figure 6. CT images of the alive rabbit at one time frame reconstructed from 29 projections. The transverse, coronal and sagittal images in (a)-(b), (c)-(d) and (e)-(f) were reconstructed using the FDK, SOTA, and our reconstruction algorithms, respectively. The images (a), (c) and (e) are volume rendering images from FDK, SOTA and our proposed methods. (b), (d) and (f) are transverse, sagittal and coronal images from FDK, SOTA and AI-empowered interior tomography methods. FDK results are unacceptable for strong artifacts, poor texture and distorted structures. The images of SOTA are over-smoothened. The images reconstructed using our method has optimal image quality in terms of texture (in the yellow circles), fidelity, and visualization (in the red and yellow circles as well as the image structure indicated by red

arrows). The display window of (b), (d) and (f) is [0 0.065] in terms of the linear attenuation coefficient.

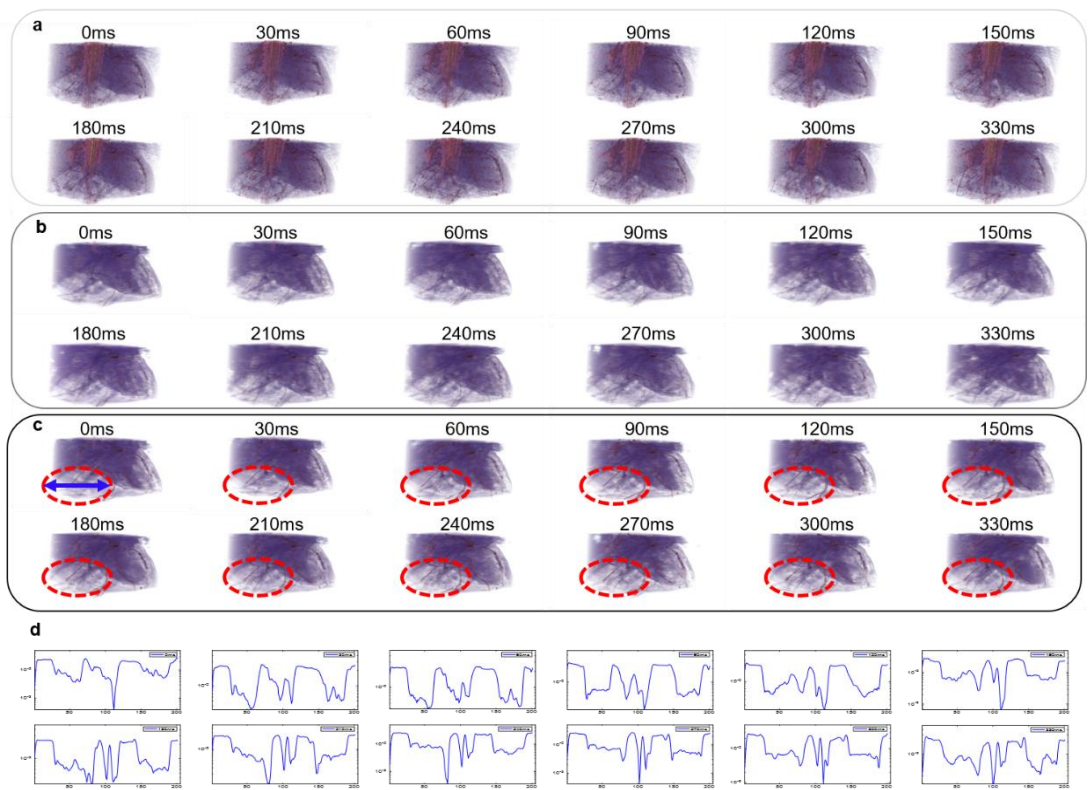

Figure 7. Sequential CT images of the alive rabbit at four time instants to visualize the real-time cardiac dynamics from only 29 projections. The images in (a), (b) and (c) were reconstructed using the FDK, SOTA and AI-empowered interior tomography methods, respectively. The images in (d) were profiles indicated by blue position in (c) with our reconstruction method. The changes of thin blood vessels can be found in (c) but they cannot be seen in the images reconstructed using FDK and SOTA methods. It clearly shows that the system resolution reaches 30ms resolution, which is the highest imaging speed of micro-CT ever reported.

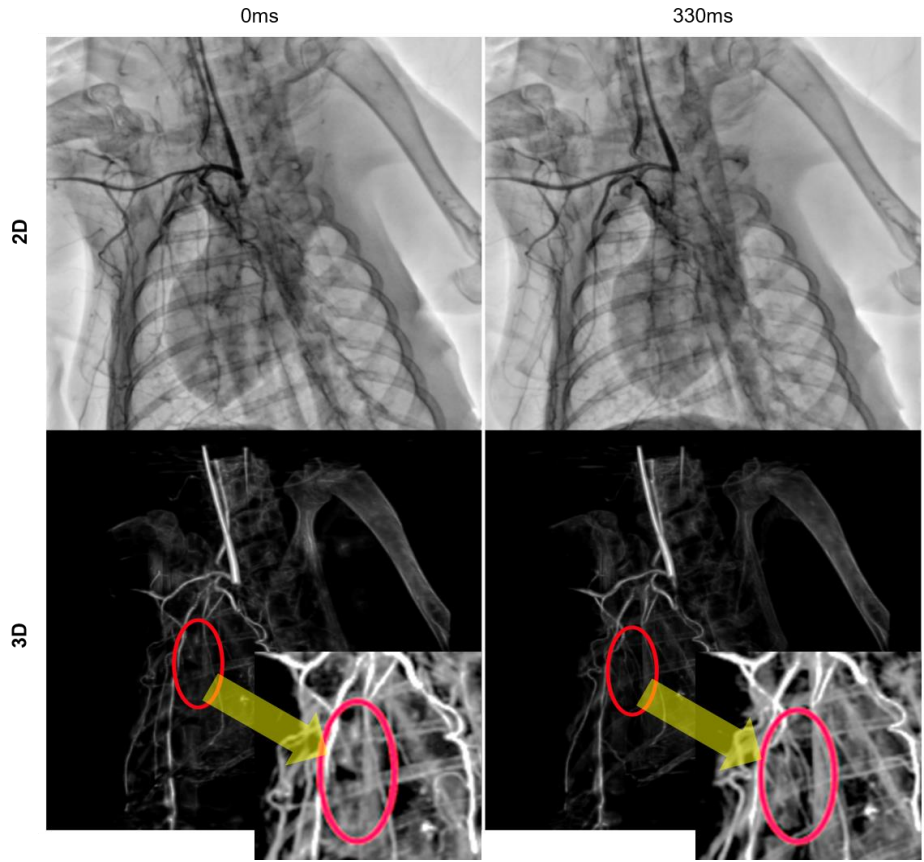

Figure 8. Dynamic CT perfusion. With 30ms temporal resolution enabled by our AI-empowered interior tomography approach, the perfusion process can be observed in real-time. The 1st row shows projections in two different time instants, and the 2nd row presents the corresponding reconstruction results respectively.

### III. Discussions & Conclusion

Our work has several major innovations. First, our SMART system enables the state-of-the-art temporal resolution through parallel acquisition of sufficiently many cone-beam projections, thereby reaching temporal resolution of 30ms. As previously described, the extremely high temporal resolution enables dynamic imaging of the chest in live rats and rabbits. Encouragingly, the reconstructed images also have high soft tissue resolution and can clearly distinguish the main organ structures in the chest tomographic images. Second, our SMART reconstruction method produces unprecedented image quality of interior CT reconstruction from only 29 views, setting the record for in the area of interior tomography. The radiation dose of CT examination is one of the main concerns of radiologists and patients. Different from conventional CT, our SMART system adopts the imaging principle similar to cone beam CT, which can greatly reduce the radiation dose while ensuring the imaging quality. This imaging method can reduce the concern of doctors and patients about the radiation dose of CT examination and expand the application of CT examination. Third, the real-time spatiotemporal tomographic imaging performance opens a door to research opportunities in not only dynamic cardiac imaging but also contrast-enhanced cancer studies. The extremely high temporal resolution can minimize the impact of heart rate on the image, and the contraindication about heart rate in the traditional cardiac CT examination will no longer exist. As shown in Figure 8, the extremely high temporal resolution can maximize the restoration of the perfusion process of the contrast agent in the tissue and improve the accuracy of perfusion-related parameters. Finally, our SMART system could be viewed as a precursor for a clinical prototype.

Our SMART system has been shown significantly improving temporal resolution to meet the real-time imaging requirement for cardiac imaging. Given the parallel-imaging hardware architecture, a major attention should be paid to scatter correction, geometric calibration, and noise reduction. As far as the image reconstruction is concerned, compared with classic priors, the advantages of the AI-empowered interior tomography method includes (1) incorporation of an advanced deep prior into tensor dictionary-based sparsified reconstruction, regularizing the solution space by combining different timeframes; (2) instantaneous reconstruction based on the current data frame effectively regularized by classic and deep image priors; and (3) superior image quality by synergizing deep network-based reconstruction and tensor dictionary learning.

As our future studies, cardiac imaging of larger animals will be performed to establish the clinical utility of the SMART architecture. It is underlined that realistic image texture and conspicuity of subtle low-contrast lesions are retained in the AI-empowered interior tomography approach images, which are clinically important [25]. If these advantages are demonstrated in large patient studies as well, numerous clinical implications will be exciting.

In conclusion, for the first time we have experimentally demonstrated the feasibility of the multi-source interior preclinical CT system for cardiac imaging in small animal models by synergizing compressed sensing and deep learning innovatively. It has been established that our AI-empowered interior tomography approach reconstructs images of ultrahigh temporal resolution and consistently produces nearly real-time reconstruction results of the beating heart in live animals. We believe that such a SMART imaging technology has a significant potential for dynamic imaging applications in preclinical, clinical and other settings.

**Methods**

**A. Compressed Sensing Inspired Reconstruction**

Solving the CT image reconstruction problem is to recover an underlying image from projection data. Let $A \in \mathbb{R}^{m \times N} (m \ll N)$ be a discrete-to-discrete linear transform representing a CT imaging system model from image pixels to detector readings; $y \in \mathbb{R}^m$ is an original dataset, $e \in \mathbb{R}^m$ is data noise in $y$, and $x \in \mathbb{R}^N$ is the image to be reconstructed, and most relevantly $m \ll N$ signifies that the inverse problem is highly under-determined. Furthermore, $L$ represents a sparsifying transform to enforce prior knowledge on the image content. Conventionally, a feasible solution can be achieved by optimizing the following $\ell_1$-norm:

$$x^* = \underset{x}{\mathrm{argmin}} \ \|Lx\|_1, \ \text{subject to } y = Ax + e. \tag{1}$$

In most cases of CT image reconstruction, the optimization problem Eq. (1) is solved using an iterative algorithm. Eq. (1) can be converted to the following minimization problem:

$$x^* = \underset{x}{\mathrm{argmin}} \frac{1}{2} \|y - Ax\|_2^2 + \lambda \|Lx\|_1, \tag{2}$$

where $\lambda > 0$ balances the data fidelity $\frac{1}{2}\|y - Ax\|_2^2$ and an image-based sparsity $\|Lx\|_1$. The goal of Eq. (2) is to find an optimized solution by minimizing the objective function. In this context, there are different regularized priors considered in the past years, including total variation [26], low-rank [27], low-dimensional manifold [28], sparse coding [29], and especially tensor-based dictionary learning [30].

A tensor is a multidimensional array. The $N^{th}$ order tensor is defined as $\mathcal{X} \in \mathcal{R}^{I_1 \times I_2 \times \ldots \times I_N}$, whose element is $x_{i_1 \times i_2 \ldots i_N}$, $1 \leq i_n \leq I_n$ and $n = 1,2,\ldots,N$. Particularly, if $N$ equals 1 or 2, the corresponding tensor is degraded to a vector or matrix. A tensor can be multiplied by a vector or a matrix. Therefore, the mode-$n$ product of a tensor $\mathcal{X}$ with a matrix $H \in \mathcal{R}^{J \times I_n}$ can be defined by $X \times_n H \in \mathcal{R}^{I_1 \times I_2 \times \ldots \times I_{n-1} \times J \times I_{n+1} \times \ldots \times I_N}$, whose element in $\mathcal{R}^{I_1 \times I_2 \times \ldots \times I_{n-1} \times J \times I_{n+1} \times \ldots \times I_N}$ is calculated as $\sum_{i_n=1}^{I_n} x_{i_1 \times i_2 \ldots i_N} h_{j \times i_n}$. In this work, we only consider the case where $\mathcal{X}$ is a 3$^{rd}$ tensor.

Suppose that there are a set of the 3$^{rd}$-order tensors $\mathcal{X}^{(t)} \in \mathcal{R}^{I_1 \times I_2 \times I_3}$ and $t = 1,2,\ldots,T$. Tensor-based dictionary learning can be implemented by solving the following optimization problem:

$$\underset{D,\alpha_t}{\operatorname{argmin}} \sum_{t=1}^{T} \left\| \mathcal{X}^{(t)} - D \times_4 \alpha_t \right\|_F^2, \text{ s.t., } \|\alpha_t\|_0 \leq L_1, \qquad (3)$$

where $D = \{D^{(k)}\} \in \mathcal{R}^{I_1 \times I_2 \times I_3 \times K}$ is a tensor dictionary, $T$ and $L_1$ represent the number of atoms in the dictionary and level of sparsity, $\|\cdot\|_F$ and $\|\cdot\|_0$ denote the Frobenius-norm and L$_0$-norm, respectively.

The K-CPD algorithm can be employed to train a tensor dictionary. The minimization problem Eq. (1) can be solved using the alternative direction minimization method (ADMM). The first stage is to update the sparse coefficient matrix using the multilinear orthogonal matching pursuit (MOMP) technique for a fixed tensor dictionary. The second stage is to update the tensor dictionary given a sparse coefficient matrix. Through alternatively updating the vector of sparse coefficients and tensor dictionary, both will be gradually optimized.

The tensor dictionary reconstruction model in cone-beam geometry can be formulated as

$$\underset{\mathcal{X},\alpha_s,m_s}{\operatorname{argmin}} \frac{1}{2}\|\mathcal{Y} - A\mathcal{X}\|_2^2 + \lambda(\sum_s \|\mathbb{Z}_s(\mathcal{X}) - D_m \times_4 m_r - D \times_4 \alpha_s\|_F^2 + \sum_s \kappa_s \|\alpha_s\|_0). \qquad (4)$$

where $\mathcal{X} \in \mathcal{R}^{I_1 \times I_2 \times I_3}$ and $\mathcal{Y} \in \mathcal{R}^{J_1 \times J_2}$ are the 3$^{rd}$-order reconstructed image and projection tensors respectively, $I_1$, $I_2$ and $I_3$ are for the reconstructed image volume, $J_1$ and $J_2$ for the numbers of detector cells and projection views respectively, $m_r$ presents the mean vector of each channel, the operator $\mathbb{Z}_s$ extracts the $s^{th}$ tensor block ($N \times N \times M$) from $\mathcal{X}$, and $\alpha_s \in \mathcal{R}^K$ is the vector of sparse representation coefficients for the $r^{th}$ tensor block. $D = \{D^{(k)}\} \in \mathcal{R}^{N \times N \times M \times K}$ is a trained tensor dictionary. $D_m = \{D_m^{(k)}\} \in \mathcal{R}^{N \times N \times M \times S}$ represents the mean removal process.

To solve the problem of Eq. (4), we introduce $\mathcal{Z}$ and convert Eq. (4) as follows:

$$\underset{\mathcal{X},\mathcal{Z},\mathcal{W},\alpha_s,m_s}{\operatorname{argmin}} \frac{1}{2}\|\mathcal{Y} - A\mathcal{X}\|_2^2 + \frac{\eta}{2}\|\mathcal{X} - \mathcal{Z} - \mathcal{W}\|_2^2 + \lambda(\sum_s \|\mathbb{Z}_s(\mathcal{Z}) - D_m \times_4 m_s - D \times_4 \alpha_s\|_F^2 +$$

$$\sum_s \kappa_s \|\alpha_s\|_0). \qquad (5)$$

where $\eta > 0$ is a balance factor. The problem Eq. (5) can be solved by dividing it into the following sub-problems:

$$\underset{\mathcal{X}}{argmin} \frac{1}{2}\|\mathcal{Y} - A\mathcal{X}\|_2^2 + \frac{\eta}{2}\|\mathcal{X} - \mathcal{Z}^{(k)} - \mathcal{W}^{(k)}\|_2^2. \tag{6}$$

$$\underset{\mathcal{Z},\alpha_s}{argmin} \frac{1}{2}\|\mathcal{X}^{(k+1)} - \mathcal{Z} - \mathcal{W}^{(k)}\|_2^2 + \lambda \left(\sum_s \|\mathbb{Z}_s(\mathcal{Z}) - D_m \times_4 m_s^{(k)} - D \times_4 \alpha_s\|_F^2 + \sum_s \kappa_s \|\alpha_s\|_0\right), \tag{7}$$

$$\underset{m_s}{argmin} \|\mathbb{Z}_s(\mathcal{Z}^{(k+1)}) - D_m \times_4 m_s - D \times_4 \alpha_s^{(k+1)}\|_F^2, \quad s = 1, \dots, S, \tag{8}$$

$$\underset{\mathcal{W}}{argmin} \frac{1}{2}\|\mathcal{X}^{(k+1)} - \mathcal{Z}^{(k+1)} - \mathcal{W}\|_2^2. \tag{9}$$

Based on Eq. (6), we compute $\mathcal{X}$ iteratively:

$$\mathcal{X}^{(k+1)} = \mathcal{X}^{(k)} - (A^T A + \eta I)^{-1} \left(A^T(A\mathcal{X}^{(k)} - y) + \eta(\mathcal{X}^{(k)} - \mathcal{Z}^{(k)} - \mathcal{W}^{(k)})\right). \tag{10}$$

Eq. (7) is a typical tensor dictionary learning problem, and can be easily solved. The solutions to Eqs. (7) and (8) can be directly obtained.

**C. SMART Network**

Given the above-described key algorithmic ingredients, we are now ready to describe our overall reconstruction scheme integrating time-frame information into sparsified prior and deep prior, which is referred to as the AI-empowered interior tomography reconstruction network. To fully understand our proposed reconstruction methodology, let us assume a continuously moving region within an object. All sources are positioned on the circular imaging ring to simultaneously radiate an object at each time frame but a dataset of 29 projections are too sparse to obtain high quality images. Also, the field-of-view of our SMART system is too small to scan a relatively large animal (such as a rabbit), interior CT reconstruction naturally becomes a challenge in this scenario.

To reconstruct a high-quality image from such an under-sampling and truncated dataset, it is instrumental to capitalize the synergy among different timeframes. One way is to incorporate a prior image to impose a constrain in the image space. The quality of a prior image will have a great impact on the final reconstruction. To obtain a high-quality prior image, we can collect sufficient data from many different timeframes to reconstruct a prior image. Because the object varies aperiodically, to the first order approximation, the precision rotation table is rotated to acquire data for different cardiac phases with inconsistent anatomical configurations. These resultant projections from different timeframes can be considered as complete (sufficiently many viewing angles) but inconsistent due to cardiac motion. Then, we pre-process the projection data and perform deep learning-based reconstruction. That is, we rearrange the projections in a chronological order for spatiotemporal sparsity-promoting image reconstruction. The overall workflow of our reconstruction approach is illustrated in Figure 1c, which can be divided into the three stages: deep group-prior prediction, sparsity group-prior reconstruction, and temporal sensing refinement.

*(1). Deep group-prior prediction*
The first stage of our proposed AI-empowered interior tomography network focuses on performing deep network-based reconstruction using the complete but inconsistent projection dataset, where the structure and intensity of different timeframes are considered. In this stage, we only need to reconstruct an initial image volume using an analytic reconstruction method (such as the FDK algorithm), which can be treated as the group-based averaged image $X$. Second, the deep interior CT reconstruction network is designed to refine the group-based reconstruction. The interior CT reconstruction network is demonstrated in Figure 1d, where the inputs includes the original analytic reconstruction results and the mask. Note that the mask $M$ is determined by the diameter of FOV, the region is covered by the FOV is set to 1 while the outside of the FOV is 0. In the interior CT network, there are two branches by performing the Hadamard product between $X$ with $M$ to reach $X_1$ and $X_2$, i.e.,

$$\begin{cases} X_1 = X \odot M \\ X_2 = X - X \odot M \end{cases}, \tag{11}$$

Next, the $X_1$ and $X_2$ pass through an encode-decode sub-network to achieve $X_3 = g_{\theta_1}(X_1)$ and $X_4 = g_{\theta_2}(X_2)$ (more details about the sub-network can be found in [31]). Finally, the output of the interior CT network is the fusion results of $X_3$ and $X_4$, which can be given as $X_5 = g_{\theta_3}(X_3, X_4)$. The fusion module consists of one convolutional layer, where both kernel and stride sizes are set to 1.

With interior tomography, the region-of-interest (ROI in the Figure 1d) can be reconstructed up to high image quality (as shown in the Figure 1d). The total loss function of the interior CT network consists of two parts:

$$Loss_{\text{total}} = \frac{1}{N}\sum_{n=1}^{N}(\|X_5^n \odot M - (X^*)^n \odot M\|_F^2 + \lambda \|X_5^n - X_5^n \odot M - (X^*)^n + (X^*)^n \odot M\|_F^2),$$
(12)

where $\lambda > 0$ represents the factor balancing the components from ROI and exterior region, $X^*$ is set to 0.1 representing the corresponding label, and N is the number of training datasets. In this stage, we reconstruct slices one-by-one.

*(2). Sparsity group-prior reconstruction*

For cardiac CT imaging, structures are deformed across different timeframes. To improve the reconstruction performance using under-sampling sparse measurements, it is feasible to reconstruct high-quality group average images from these group projections using sparsity regularization techniques. We reconstruct a high-quality group-based image using the following model

$$\underset{\mathcal{X},\alpha_{s_1},m_{s_1},\alpha_{s_2},m_{s_2}}{argmin} \frac{1}{2}\|\mathcal{Y}' - A\mathcal{X}\|_2^2 + \lambda_1 \left(\sum_{s_1}\|\mathbb{Z}_{s_1}(\mathcal{X}) - D_m \times_4 m_{s_1} - D \times_4 \alpha_{s_1}\|_F^2 + \sum_{s_1}\kappa_{s_1}\|\alpha_{s_1}\|_0\right) +$$

$$\lambda_2 \left(\sum_{s_2}\|\mathbb{Z}_{s_2}(\mathcal{X} - \mathcal{X}^D) - D_m \times_4 m_{s_2} - D \times_4 \alpha_{s_2}\|_F^2 + \sum_{s_2}\kappa_{s_2}\|\alpha_{s_2}\|_0\right), \tag{13}$$

where $\mathcal{Y}'$ is the combined projections with different time-frame measurements and $\lambda_1$ is a balance factor to trade-off the data fidelity term and regularization term. To obtain the solution of Eq. (13), a similar strategy for solving Eq. (6) is employed. Here, we introduce two $\mathcal{Z}_1$ and $\mathcal{Z}_2$ to replace with $\mathcal{X}$ and $\mathcal{X} - \mathcal{X}^D$, and $\mathcal{X}^D$ denotes the prior image obtained in the deep group-prior prediction stage. Hence, Eq. (6) can be converted into

$$\underset{\mathcal{X}}{argmin} \frac{1}{2}\|\mathcal{Y} - A\mathcal{X}\|_2^2 + \frac{\eta_1}{2}\|\mathcal{X} - \mathcal{Z}_1^{(k)} - \mathcal{W}_1^{(k)}\|_2^2 + \frac{\eta_2}{2}\|\mathcal{X} - \mathcal{Z}_2^{(k)} - \mathcal{W}_2^{(k)}\|_2^2. \tag{14}$$

where $\eta_1 > 0$ and $\eta_2 > 0$ need to be empirically chosen. Similar to what we described above, $\mathcal{W}_1$ and $\mathcal{W}_2$ are error feed-back variables (put the explanation earlier and cite a paper) to be updated. Regarding the tensor dictionary learning, how to obtain high-quality dictionary plays an important role in controlling image quality. Here, the prediction in the first stage was employed to train the tensor dictionary. It is beneficial to further explore deep prior using tensor dictionary regularization and provide a high-quality tensor dictionary based on group-based prior images.

*(3). Temporal sensing refinement*

Till now, we can reconstruct high-quality group images from different timeframes. However, the reconstructed images may include blurred and other artifacts induced by dynamically changing structures. To reconstruct small structural changes, it is feasible to combine temporal measurement data and group prior images into the unified reconstruction model. To improve the temporal resolution of SMART system, the reconstructed results in the second stage are used as the prior images in the iteration reconstruction model of Eq. (13). Then, only the original timeframe measurement is used to refine the final results, according to the same tensor dictionary used in the 2$^{nd}$ stage.

**D. Experimental Design**

*(1). Setup, Data and Codes*

To validate the feasibility of our SMART system and AI-empowered interior tomography network for dynamic interior tomography, we performed pre-clinical experiments with the SMART system and produced encouraging results. Several preclinical datasets were collected from dead and alive rats as well as alive rats. Experimental animal studies were performed under the Animal Research: Reporting of In Vivo Experiments (ARRIVE) guidelines. Five adult male animals of 250-300 grams were purchased from the Jie Si Jie Laboratory Animal Co., Ltd. (Shanghai, China). The animal experimental protocol was approved by the Institutional Animal Care and Use Committee (IACUC) of Shanghai Jiao Tong University, Shanghai, China.

Since the original scans are in cone-beam geometry, we need to reconstruct multiple timeframes to observe the dynamic process. For our system prototype, the source-to-detector distances of the 29 source-detector pairs are between 2,016mm and 2,087mm. The source-to-isocenter distances of the 29 source-detector pairs are between 1,079mm and 1,167mm. The plat detector contains 768×768 cells, each of which covers an area of 0.2 × 0.2 mm$^2$. As a result, the diameter of FOV is about 81mm. There are 29 source-detector pairs simultaneously activated in every scan. Hence, 29 cone-beam projections are distributed over a full-scan angular range.

To collect the datasets used in the 1$^{st}$ stage to train the deep neural networks 10 dead rats were scanned, the procedure of training is shown in Figure 1e of the main text. Because these dead rats are small (smaller than the diameter of FOV), our system can collect the measurements without data truncation. To simulate the data truncation, the datasets from the detector panel was cropped into 450×768 cells. Thus, the two ends along the horizontal direction of the detector were throwed away. In total, 68,544 images of 512×512 pixels were extracted from 9 dead rats for training. 512 slices from the last one was used for testing. In our evaluation, the ground truth of 10 died rats were reconstructed using the 348-view tensor dictionary learning-based (TDL) algorithm rather than FDK method, as shown in Eq. (10). Because there is noise from scatter and dose instability, the TDL is good for improving the reconstructed images quality. The testing datasets are divided into two datasets from alive rats and the other from rabbits respectively.

To highlight the advantages of our reconstruction approach over the traditional algorithms, the total-variation based minimization [32] was implemented as the state-of-the-art (SOTA) compressed sensing based reconstruction method. Besides, the classical FDK method was also selected for comparison. In this study, all the source codes for deep learning reconstruction were written in Python with the Pytorch library on a NVIDA RTX3080 card. The 2$^{nd}$ and 3$^{rd}$ stages were programmed in Matlab 2021 on the Graphics Processing Unit (GPU). All programs were executed on a PC (24 CPUs @3.70GHz, 32.0GB RAM) with Windows 10.

*(2). Network Training*

The Adam method was employed to optimize all of the networks [33]. To address the inconsistency in sizes of feature maps and that of the input, we padded zeros around the boundaries before convolution. The batch size was set to 1. The number of epochs was set to 50. The learning rate was set to 2.5×10$^{-4}$ and decreased by 0.8 after each of 5 epochs.

*(3). AI-empowered Interior Tomography Approach Parameters*

Our AI-empowered interior tomography network belongs to the category of hybrid reconstruction methods since it combines deep learning, compressed sensing and algebraic iteration, which means that there are regularization parameters to be chosen in a task-specific fashion. There are at least five parameters in the 2$^{nd}$ and 3$^{rd}$ stages, the balance factors $\eta_1$ and $\eta_2$, the number of dictionary atoms *M*, the level of sparsity *L*. In our AI-empowered interior tomography network, $\eta_1$ and $\eta_2$ represent the coupling factors to balance the associated components in the 2$^{nd}$ stage, $\eta_3$ and $\eta_4$ represent the coupling factors to balance the associated components in the 3$^{rd}$ stage. $M_1$ and $M_2$ represent the number of dictionary atoms in the 2$^{nd}$ and 3$^{rd}$ stages. $L_1$ and $L_2$ represent the level of sparsity in the 2$^{nd}$ and 3$^{rd}$ stages. The specific parameters values are summarized in Table I.

Table I. AI-empowered interior tomography network parameters for dynamic cardiac imaging on the SMART system.

| Parameters | $\eta_1$ | $\eta_2$ | $\eta_3$ | $\eta_4$ | $L_1$ | $L_2$ | $M_1$ | $M_2$ |
|---|---|---|---|---|---|---|---|---|
| Dead Rat | 0.15 | 0.15 | 0.15 | 0.15 | 0.001 | 0.0004 | 5 | 5 |
| Alive Rat | 0.15 | 0.15 | 0.15 | 0.15 | 0.001 | 0.0004 | 5 | 5 |
| Rabbit | 0.15 | 0.25 | 0.30 | 0.30 | 0.0012 | 0.002 | 10 | 10 |